\documentclass[epjCONF,columns]{svjour} 
\usepackage{graphics}
\usepackage[varg]{txfonts} 
\usepackage[latin1]{inputenc}
\newcommand{\dblh}{\Phi^{++}}
\newcommand{\xsecbr}{\sigma_\phi\cdot B_{\tau\tau}}

\session-title{Hadron Collider Physics symposium, Paris 2011}
\begin{document}
\title{Recent results on beyond the standard model Higgs boson searches from CMS}
\author{ Alexander A. Savin\fnmsep\thanks{\email{asavin@mail.cern.ch}} on behalf of the CMS Collaboration }
\institute{University of Wisconsin-Madison, 1150 University Ave., Madison WI 53706-1390}
\abstract{
Two extensions of the standard model, one that includes
the seesaw mechanism of type II, and the minimal supersymmetric extention
to the standard model, are studied using up to 1.6 fb$^{-1}$ of 
data collected in proton-proton collisions at $\sqrt{s}$=7 TeV
 with the CMS detector at the LHC.}
\maketitle
\section{Introduction}
\label{intro}

An  observation of 
the doubly charged component $\Phi^{++}$ (here and below charge conjugate modes are 
implicitly included) of the triplet scalar field predicted in the 
the minimal seesaw model of type II,
would establish such mechanism in the most promising 
framework giving mass to neutrinos.
This particle carries double electric charge and decays to the same charged 
lepton pairs $\ell^+_i\ell^+_j$ allowing also lepton flavor violating decays. 
The  $\Phi^{++}$ Yukawa coupling matrix 
is proportional to the light neutrino
 mass matrix and allows to test the neutrino mass mechanism by measuring
the  branching fractions $\Phi^{++}\to \ell_i\ell_j$~\cite{HIG-11-007}.

The minimal supersymmetric extension to the standard model
(MSSM) requires the presence of two Higgs doublets.  This leads to a
more complicated scalar sector, with five massive Higgs bosons: a
light neutral CP-even state ($h$), two charged states ($H^\pm$), a heavy
neutral CP-even state ($H$) and a neutral CP-odd state ($A$).

If the charged Higgs boson mass, $m_{H^{+}}$, is smaller than the top quark mass,
the top quark can decay via
$t\rightarrow H^{+} b$ (and its charge conjugate).
The lower limit on the
charged Higgs boson mass is set to about 80~GeV/$c^2$ by LEP experiments.
For values of $\tan\beta$, the ratio of the vacuum expectation values of the 
two Higgs boson doublets, larger than 20, the charged
Higgs boson preferentially decays to $\tau$ lepton and neutrino, 
$H^{+} \rightarrow \tau ^{+} \nu _{\tau}$.
The presence of the $t\rightarrow H^{+} b$, $H^{+} \rightarrow \tau ^{+} \nu_{\tau}$ 
decay modes alters the standard model (SM)
 prediction of the $\tau$ lepton yield in the decay products of the $t\bar{t}$ pairs. 
The current upper limit on the branching
fraction $BR(t\rightarrow H^+b)$ $\simeq 0.2$ is set by the 
CDF
 and D0
experiments at the Tevatron for $m_{H^{+}}$ between 80 and 155 GeV/$c^2$, assuming $BR(H^{+} \rightarrow \tau^{+} \nu)$=1.
The dominant source of top quarks at LHC is $pp \rightarrow t \bar{t}$ process, 
therefore the charged Higgs boson
is searched for
in the subsequent decay products of the top quark pairs: 
$t \bar{t} \rightarrow H^{\pm} W^{\mp} b \bar{b}$ and
$t \bar{t} \rightarrow H^{\pm} H^{\mp} b \bar{b}$ when 
$H^{\pm}$ decays into $\tau$ lepton and neutrino~\cite{HIG-11-008}.

The mass relations among the neutral MSSM Higgs bosons are such that
if $m_A \lesssim 130$ GeV/$c^2$, at large values of the parameter
$\tan\beta$ the masses of the $h$ and $A$ are nearly degenerate, while
that of the $H$ is approximately 130 GeV/$c^2$.  If $m_A \gtrsim 130$ GeV/$c^2$,
then the masses of the $A$ and $H$ are nearly degenerate, while that
of the $h$ remains near 130 GeV/$c^2$. The precise value of the crossover
point depends predominantly on the nature of the mass mixing in the
top-squark states.
The neutral MSSM Higgs boson production is studied in its decay into
pair of tau leptons, $H \to \tau\tau$, 
in three final  states, when $\tau$ decay leptonically, one to $\mu$ and other
to $e$, $e\mu$, and when one of the $\tau$ decays hadronically, $\mu \tau_{h}$
and $e \tau_{h}$~\cite{HIG-11-020}.

\section{Doubly charge Higgs boson production}
\label{sec:1}
Both, the pair production process 
$pp \to \Phi^{++}\Phi^{--}\to \ell_i^+\ell_j^+\ell_k^-\ell_l^-$,
as well as  the associated production 
process $pp\to \Phi^{++}\Phi^{-}\to  \ell_i^+\ell_j^+\ell_k^-\nu_l$
 are studied, assuming that the $\dblh$ and $\Phi^{+}$ are degenerate in mass. 
A search is performed for an excess of events in all possible flavour combinations of the 
same charge lepton pairs 
coming from
the decays $\Phi^{++}\to \ell^+_i\ell^+_j$ without making
assumptions on the $\Phi^{++}$ branching fractions.
Both the three and four charged lepton final states are considered including at most one and two $\tau$ leptons, respectively.
The $\Phi^{++}\to W^{+}W^{+}$ decays are assumed to be suppressed.

\begin{figure}
\rotatebox{-90}{\resizebox{0.75\columnwidth}{!}{%
\includegraphics{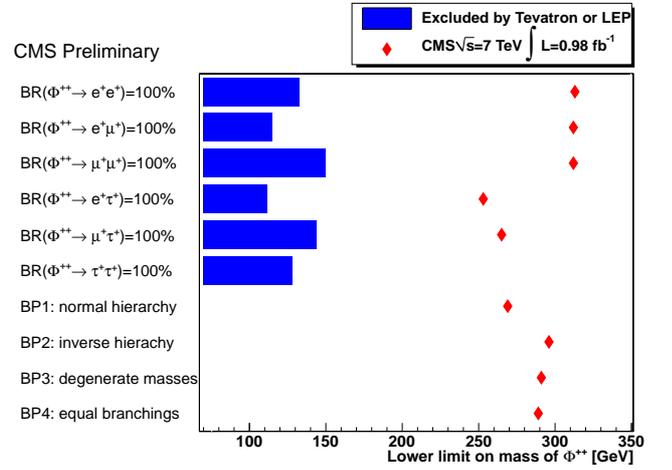}}}
\caption{Observed $\dblh$ mass limits at 95\% CL in different lepton final states. The branching fractions that are assumed in the limit calculation are indicated.}
\label{fig:punchline}
\end{figure}

In addition to the model independent search, the type II seesaw model is tested in four 
benchmark points (BP) that characterize different characteristic
neutrino mass matrix structures. BP1 describes the neutrino sector with normal mass 
hierarchy and a massless lightest neutrino.
BP2 describes the same, but with the inverse mass hierarchy. BP3 represents a degenerate 
neutrino mass spectrum with $m_1=0.2$~eV. Those three benchmark points are the extremes
allowed by varying the neutrino mass and hierarchy in the allowed ranges without 
consideration for $\theta_{13}$ or CP phases and cover a large region of the parameter space. 
The fourth benchmark point
 BP4 represents the case in which all $\Phi^{++}$ branching fractions are equal. 

The presented results are based on data corresponding to an
integrated luminosity of 0.98 $fb^{-1}$. The final states are required to have four or
three leptons with transverse momenta, $p_{\rm{T}}$, above 5 GeV/$c$ for muons and above
15 GeV/$c$ for electrons and $\tau_{\rm{h}}$. There should be at least two leptons with 
$p_{\rm{T}} > 35$ and 10 GeV/$c$ in the event.
 The $\tau_{\rm{h}}$ are reconstructed
using ``hadron plus strip''(HPS) algorithm~\cite{TAU-11-001}
that
is designed to optimize the performance of
$\tau_{\rm{h}}$ identification and reconstruction
by considering specific $\tau_{\rm{h}}$
decay modes. The algorithm provides high $\tau_{\rm{h}}$ identification efficiency,
approximately 50\% for the range of $\tau_{\rm{h}}$ energies relevant for this analysis,
while keeping the misidentification rate for jets at the level
of $\approx 1\%$ that is factor of three to four times lower than for the
algorithms used in the CMS physics technical design report.
Detailed description of the algorithm and its performance
can be found in~\cite{TAU-11-001}.

A CLs method is used for the upper limit calculations.
The results of the exclusion limit calculations are reported 
in Fig.~\ref{fig:punchline}.
As can be seen, new significantly higher limits are set in comparison to the previous LEP and 
 Tevatron bounds. 
The first limits on four benchmark points probe a
 large region of the parameter
 space of type II seesaw models.

\section{Study of minimal supersymmetric extension to the standard model}
\label{sec:2}

\subsection{Charged Higgs boson}
\label{sec:21}

The analysis is performed in three final states, $e\mu$, $\mu \tau_{\rm{h}}$ and
$\tau_{\rm{h}} \tau_{\rm{h}}$, in presence of 
large missing transverse energy 
and one b-tagged jet. Transverse momenta of all leptons are required to exceed 20 GeV/$c$,
 40 GeV/$c$ in the
$\tau_{\rm{h}} \tau_{\rm{h}}$ case.

\begin{figure}
\begin{center}
\rotatebox{0}{\resizebox{0.95\columnwidth}{!}{%
\includegraphics{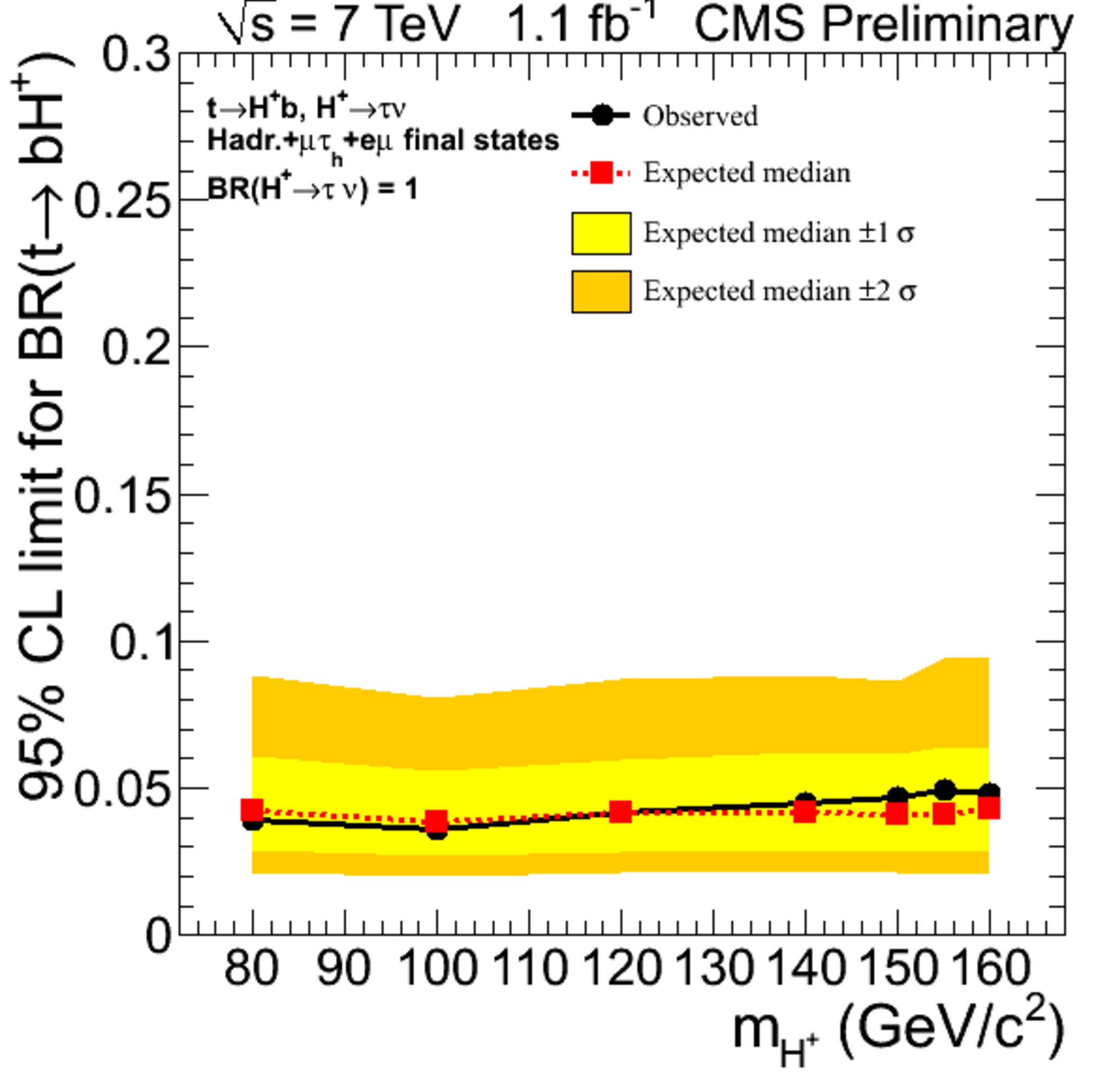}}} 
\caption{Upper limit on $BR(t\rightarrow H^{+}b)$ assuming $BR(H^{+} \rightarrow \tau \nu)$=1 
as a function
of $m_{H^{+}}$for the combination of all final states.
The yellow bands show the one- and two-standard-deviation ranges around the expected limit.}
\label{fig:limitcombination_fig67}
\end{center}
\end{figure}

A CLs method is used in order to obtain the upper limit at 95\%~CL on the excess (lack) of the
events in comparison with the expected contributions from SM.
The background and signal uncertainties
are modeled with a log-normal probability distribution 
function and their
correlations
are taken into account.
Assuming that the excess (lack) of events is due to the 
$t \rightarrow bH^{+}, ~ H^{+} \rightarrow \tau^{+} \nu$ decays, 
the 
limit is obtained
for each individual analysis, and also combined for all final states.
Figure~\ref{fig:limitcombination_fig67} shows the upper limit on 
$BR(t\rightarrow H^{+}b)$
assuming $BR(H^{+} \rightarrow \tau \nu)$=1 as a function 
of the higgs mass for the combination of all final states.
The upper limit is set 
to 4-5 \% 
for the Higgs boson mass
interval 80 $< m_{H^{+}} <$ 160 GeV/$c^2$.

\begin{figure}
\begin{center}
\rotatebox{0}{\resizebox{0.95\columnwidth}{!}{%
\includegraphics{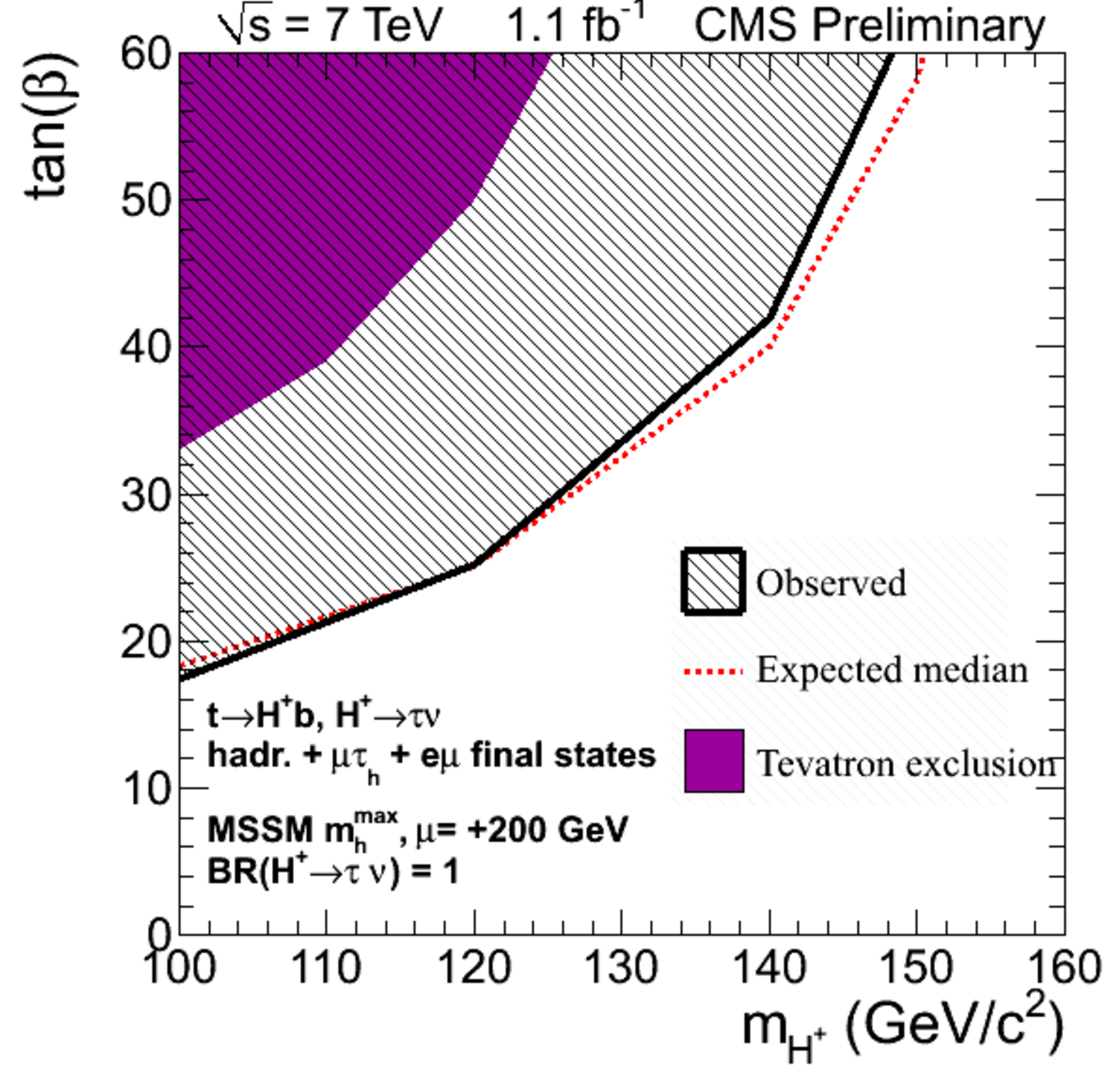}}}
\caption{The exclusion region in the MSSM $M_{H^{+}}$-tan$\beta$ parameter space obtained from
         the combined analysis for the MSSM $m_{h}^{max}$ scenario.}
\label{fig:limitcombination_fig89}
\end{center}
\end{figure}

Figure~\ref{fig:limitcombination_fig89} shows the exclusion region in the MSSM $M_{H^{+}}$-tan$\beta$
parameter space obtained from the combined analysis for the 
MSSM $m_{h}^{max}$ scenario:
$M_{SUSY} = 1$~TeV/$c^2$, $\mu = 200$~GeV/$c^2$, $M_{2} = 200$~GeV/$c^2$, 
$m_{\tilde{g}} = 0.8 \cdot M_{SUSY}$, $X_{t} = 2 \cdot M_{SUSY}$ (FD calculation),
$X_{t}^{\bar{MS}} = \sqrt{6} \cdot M_{SUSY}$ (RG calculation), 
$A_{b} = A_{t}$.  
$M_{\rm{SUSY}}$ denotes the common soft-SUSY-breaking squark mass of the third
generation; $X_t = A_t - \mu/\tan\beta$ the stop mixing parameter;
$A_t$ and $A_b$ the stop and sbottom trilinear couplings,
respectively; $\mu$ the Higgsino mass parameter; $m_{\tilde{g}}$ the
gluino mass; and $M_2$ the SU(2)-gaugino mass parameter. The value of
$M_1$ is fixed via the GUT relation $M_1 =(5/3)M_2\sin\theta_{\rm
W}/\cos\theta_{\rm W}$.  In determining these bounds on $\tan\beta$,
we have
used the central values of the Higgs boson cross sections as a
function of $\tan\beta$ reported by the LHC Higgs Cross Section
Working Group.  The cross sections have been obtained
from the {\sc GGH@NNLO} and
{\sc HIGLU} programs for the gluon-fusion process.
For the $b\bar{b}\to\phi$ process, the 4-flavor
calculation and the 5-flavor
calculation as implemented in the {\sc
BBH@NNLO} program have been combined using the
Santander scheme. Rescaling of the corresponding
Yukawa couplings by the MSSM factors calculated with
FeynHiggs has been applied.  We do not quote
limits above $\tan\beta = 60$ as the theoretical relation between
cross section and $\tan\beta$ becomes unreliable.

\subsection{Neutral Higgs boson}
\label{sec:22}

The MSSM neutral Higgs boson analysis is performed in three final states $e \mu$, $\mu \tau_{\rm{h}}$, and
$e \tau_{\rm{h}}$ with $p_{\rm{T}} > 15$ GeV/$c$ for muons, and $p_{\rm{T}} > 20$ GeV/$c$ for
$e$ and $\tau_{\rm{h}}$. In $e\mu$ final states one of the lepton is required to have
$p_{\rm{T}} > 20$ GeV/$c$ and other $p_{\rm{T}} > 10$ GeV/$c$.
Each final state is subdivided into two categories, one with no b-tagged jets with 
$p_{\rm{T}} > 20$ GeV/$c$ and the other with at least one b-tagged jet. 

The largest source of events selected with these requirements comes from
the SM $Z \to \tau\tau$ production.
The contribution from this process is estimated
using a sample of simulated events, 
normalized to 
the number of observed
$Z \to \mu\mu$ and $Z \to ee$ events in data.
A significant source of background arises from QCD multijet events and
$W$+jets events in which a jet is misidentified as $\tau_{\rm{h}}$, and
there is a real or misidentified $e$ or $\mu$. The rates for these
processes are estimated using the number of observed 
events, where both reconstructed leptons have the same charge, same-sign combination.
Other background processes include $t\bar{t}$ production
and $Z \to ee/\mu\mu$ events, particularly in the $e\tau_{\rm{h}}$
channel, due to the 2--3\% probability for electrons to be
misidentified as $\tau_{\rm{h}}$~\cite{TAU-11-001}. The small
fake-lepton background from $W +$jets and QCD for the $e\mu$
channel is estimated using data.

The event generator {\sc PYTHIA}, and {\sc POWHEG}, and {\sc MadGraph} are
used to model the Higgs boson signal and other backgrounds.  The 
{\sc TAUOLA} package is used for tau decays in all cases.

To distinguish the Higgs boson signal from the background, we
reconstruct the visible mass, defined as the invariant mass of
the visible tau decay products. The observed
visible mass distributions are fit in each case to the sum of the SM
backgrounds and the Higgs boson signal.
The visible mass spectra show no evidence for the presence of a Higgs boson
signal, and are used to set a 95\% CL upper bound on the
product of the Higgs boson cross section and the tau-pair branching fraction, 
 $\xsecbr$. 
\begin{figure}
\begin{center}
\rotatebox{0}{\resizebox{0.95\columnwidth}{!}{%
\includegraphics{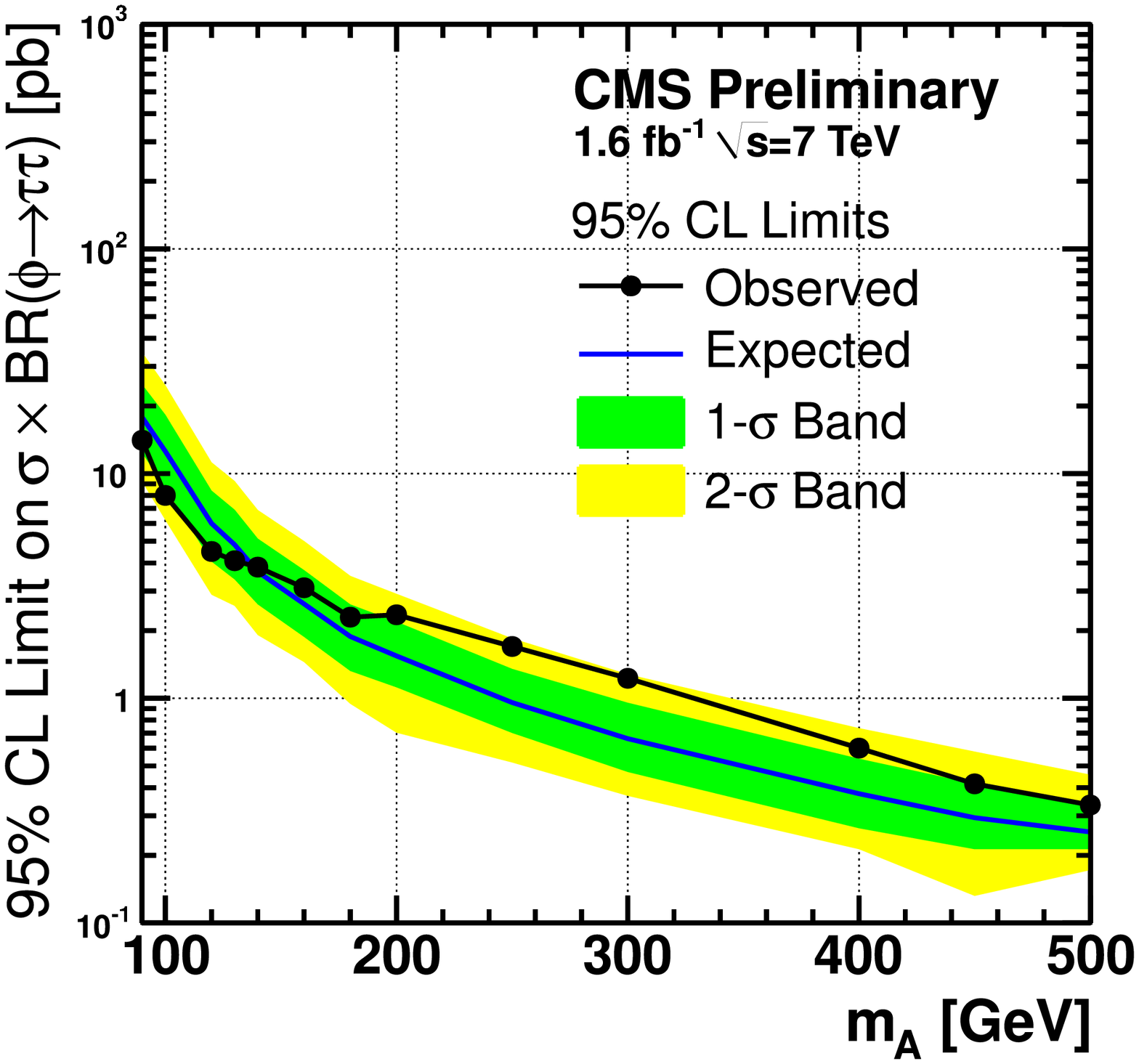}}}
    \caption{The expected one- and two-standard-deviation ranges and observed 95\%  CL upper limits on
    $\xsecbr$ as a function of $m_A$.  The signal acceptance is based on the MSSM model described in
    the text, assuming $\tan\beta=30$.}
    \label{xsec-limit}
\end{center}
\end{figure}

Figure~\ref{xsec-limit} shows the upper bound on $\xsecbr$ as
a function of $m_A$, where we use as the signal acceptance model the
combined visible mass spectra from the $gg$ and $b\bar{b}$
production processes for $h$, $A$, and $H$, and assuming
$\tan\beta=30$.  The plot also shows the one- and
two-standard-deviation range of expected upper limits for various
potential experimental outcomes.  The observed limits are well within
the expected range assuming no signal.

\begin{figure}
\begin{center}
\rotatebox{0}{\resizebox{1.0\columnwidth}{!}{%
\includegraphics{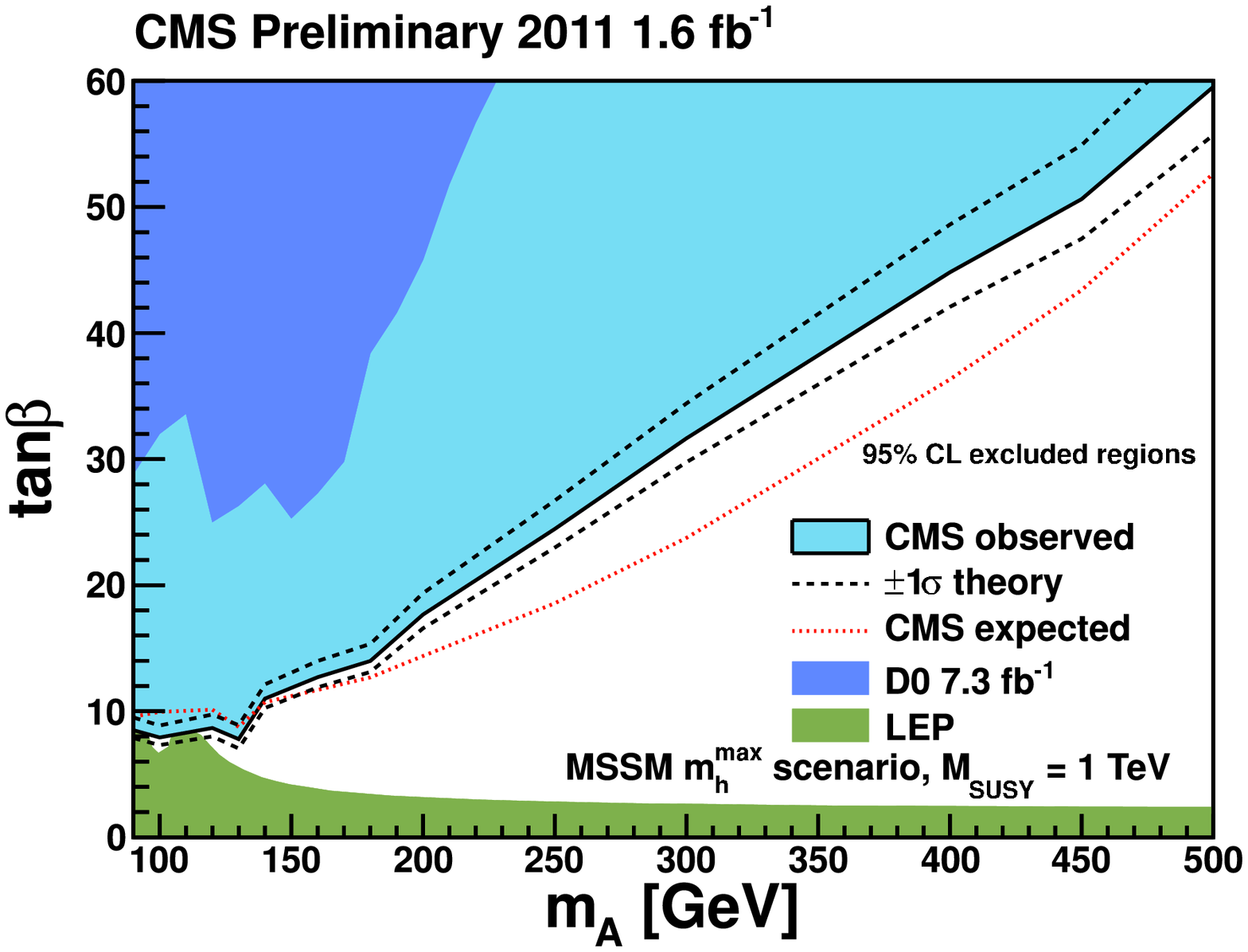}}}
    \caption{Region in the parameter space of $\tan\beta$ versus $m_A$ excluded
             at 95\% CL in the context of the MSSM $m^{\rm max}_h$ scenario, with
             the effect of $\pm 1\sigma$ theoretical uncertainties shown.  The
             other shaded regions show the 95\% CL excluded regions from the
             LEP and Tevatron experiments.}
    \label{tanbeta-ma}
\end{center}
\end{figure}

One can interpret the upper limits on $\xsecbr$ in the MSSM parameter
space of $\tan\beta$ versus $m_A$ for an example scenario in the
same way it was done above for the charged Higgs. The results are shown
in Fig.~\ref{tanbeta-ma}.  
They exclude a region in $\tan\beta$ down to values
smaller than those excluded by the Tevatron
experiments for $m_A\lesssim$ 140~GeV/$c^2$, and
significantly extend the excluded region of MSSM parameter space at
larger values of $m_A$.  Figure~\ref{tanbeta-ma} also shows the region
excluded by the LEP experiments.


\begin{thebibliography}{}
\bibitem{HIG-11-007}
CMS Collaboration, CMS-PAS-HIG-11-007, 2011, and references therein
\bibitem{HIG-11-008}
CMS Collaboration, CMS-PAS-HIG-11-008, 2011, and references therein
\bibitem{HIG-11-020}
CMS Collaboration, CMS-PAS-HIG-11-020, 2011, and references therein
\bibitem{TAU-11-001}
CMS Collaboration, CMS-PAS-TAU-11-001, 2011, accepted by JINST
\end{thebibliography}
\end{document}